\documentclass[
aps,
prb,
superscriptaddress,
citeautoscript,
amsmath,
amssymb,
amsfonts,
showkeys,
floatfix,
reprint,
]{revtex4-2}

\usepackage{iftex}
\iftutex                        
  \usepackage{fontspec}
  \usepackage{polyglossia}
  \setmainlanguage[variant=american]{english}
\else                           
  \usepackage[T1]{fontenc}      
  \usepackage[utf8]{inputenc}   
  \usepackage[english]{babel}   
\fi

\usepackage{xparse}
\usepackage{ifthen}
\newboolean{draft}

\NewDocumentCommand\change{om}{%
  \ifthenelse{\boolean{draft}}
  {\IfNoValueTF{#1}{{\color{orange}#2}}{{\color{lightgray}#1}/{\color{orange}#2}}}
  {#2}%
}

\usepackage{xcolor}

\usepackage{graphicx}
\graphicspath{ {.} }

\usepackage{microtype}
\usepackage{csquotes}

\usepackage{array}
\usepackage{dcolumn}
\usepackage{makecell}
\usepackage{booktabs}
\usepackage{multirow}
\newcolumntype{d}[1]{D{.}{.}{#1}} 

\usepackage{amsmath}
\usepackage{mathtools}
\usepackage{amssymb}
\usepackage{amsfonts}
\usepackage{bm}
\usepackage{derivative}

\usepackage{natbib}

\usepackage{siunitx}
\DeclareSIUnit{\angstrom}{\text{\AA}}

\usepackage{hyperref}
\hypersetup{
  plainpages=false,
  unicode=true,
  colorlinks=true,
  linkcolor=blue,
  citecolor=blue,
  urlcolor=blue,
}
\usepackage[all]{hypcap}

\def\mydate{\leavevmode\hbox{\the\year-\twodigits\month-\twodigits\day}}
\def\twodigits#1{\ifnum#1<10 0\fi\the#1}

\DeclarePairedDelimiter{\parens}{\lparen}{\rparen} 
\DeclarePairedDelimiter{\cbra}{\{}{\}}             
\DeclarePairedDelimiter{\abra}{\langle}{\rangle}   
\DeclarePairedDelimiter{\vbra}{\lvert}{\rvert}     

\DeclarePairedDelimiter{\ket}{\lvert}{\rangle} 
\DeclarePairedDelimiterX{\braket}[2]{\langle}{\rangle}{#1\,\delimsize\vert\,\mathopen{}#2} 

\newcommand*{\iu}{\ensuremath{\mathrm{i}}}         
\newcommand*{\en}{\ensuremath{\mathrm{e}}}         

\newcommand*{\abs}[1]{\vbra{#1}}   

\newcommand*{\set}[1]{\cbra{#1}}     

\renewcommand*{\vec}[1]{\ensuremath{\boldsymbol{#1}}} 
\newcommand*{\func}[2]{\ensuremath{#1\parens{#2}}}    

\newcommand*{\sdown}[1]{\ensuremath{\bar{#1}}}

\newcommand*{\pgroup}[2]{\ensuremath{{#1}_{\mathrm{#2}}}}

\newcommand*{\Ci}{\ensuremath{\mathrm{C_{i}}}}
\newcommand*{\Oi}{\ensuremath{\mathrm{O_{i}}}}
\newcommand*{\CiOi}{\ensuremath{\mathrm{C_{i}O_{i}}}}

\newcommand*{\dSCF}{\ensuremath{\Delta \mathrm{SCF}}}

\newrobustcmd\NoteMark[1]{\textsuperscript{#1}}

\setboolean{draft}{false}

\begin{document}

\title[C-center]{Optical lineshapes of the C-center in silicon from \textit{ab
    initio} calculations: Interplay of localized modes and bulk phonons}

\author{Rokas~Silkinis}
\email{rokas.silkinis@ftmc.lt}
\affiliation{Department of Fundamental Research, Center for Physical Sciences and Technology (FTMC), Vilnius LT--10257, Lithuania}

\author{Marek~Maciaszek}
\affiliation{Department of Fundamental Research, Center for Physical Sciences and Technology (FTMC), Vilnius LT--10257, Lithuania}
\affiliation{Faculty of Physics, Warsaw University of Technology, Koszykowa 75, 00-662 Warsaw, Poland}

\author{Vytautas~Žalandauskas}
\affiliation{Department of Fundamental Research, Center for Physical Sciences and Technology (FTMC), Vilnius LT--10257, Lithuania}

\author{Marianne~Etzelmüller~Bathen}
\author{Lasse~Vines}
\affiliation{Department of Physics/Centre for Materials Science and Nanotechnology, University of Oslo, P.O. Box 1048, Blindern, Oslo N-0316, Norway}

\author{Audrius~Alkauskas}
\thanks{Deceased}
\affiliation{Department of Fundamental Research, Center for Physical Sciences and Technology (FTMC), Vilnius LT--10257, Lithuania}

\author{Lukas~Razinkovas}
\email{lukas.razinkovas@ftmc.lt}
\affiliation{Department of Fundamental Research, Center for Physical Sciences and Technology (FTMC), Vilnius LT--10257, Lithuania}
\affiliation{Department of Physics/Centre for Materials Science and Nanotechnology, University of Oslo, P.O. Box 1048, Blindern, Oslo N-0316, Norway}




\begin{abstract}
  In this work, we present a first-principles density functional
  theory (DFT) computational investigation of the luminescence and
  absorption lineshapes associated with the neutral carbon-oxygen
  interstitial pair (\CiOi) defect in silicon. We obtain the
  lineshapes of the defect in the dilute limit using a computational
  methodology that constructs dynamical matrices of supercells
  containing tens of thousands of atoms, utilizing systems directly
  accessible through DFT. Both perturbed bulk phonons and localized
  vibrations contribute to the phonon sideband. We achieve excellent
  agreement with experimental luminescence data. Our findings further
  reinforce the attribution of the well-known C-line in silicon to the
  neutral \CiOi\ complex.
\end{abstract}


\maketitle

\section{Introduction\label{sec:introduction}}

In recent years, quantum technologies have gained significant
attention in the scientific community due to their potential
applications in quantum
sensing~\cite{2017--Degen--RMP--QS,2013--Le-Sage--N--OMIL,2014--Rondin--RPP--MNVD,2022--Vindolet--PRB--OPSG},
quantum
communication~\cite{2007--Gisin--NP--QC,2023--Al-Juboori--AQT--QKDU},
and quantum computing~\cite{2010--Ladd--N--QC,2010--Weber--PNAS--QCD}.
Silicon, with its established commercial success and role in
solid-state devices, presents an intriguing host material for the
development of future quantum technologies, with a variety of
optically active point defects. These defects are of particular
interest as single-photon emitters that could potentially be used for
short- and long-distance exchange of information. Recent
demonstrations of single photon emission from the G-center defect in
silicon~\cite{2020--Hollenbach--OE--ETSP,2020--Redjem--NE--SAAS,2021--Durand--PRL--BDNI}
has resulted in revisiting previously studied silicon defects, such as
the W-~\cite{2020--Buckley--OE--OPWC,2022--Baron--AP--DSWC} and
T-centers~\cite{2022--Dhaliah--PRM--FPST}. Due to silicon's relatively
small \qty{1.1}{\eV} band gap, all of these defects are active in the
technologically relevant near-infrared region of light.

Carbon and oxygen atoms are among the most frequent sources of
impurities in silicon. Substitutional carbon can capture silicon
self-interstitials, generated by particle irradiation, and form carbon
interstitials (\Ci)~\cite{1996--Zhao--MP--IDRS}. The \Ci\ is mobile at
room
temperature~\cite{1976--Noonan--JAP--PSIE,1989--Kimerling--MSF--IDRS}
and may capture an interstitial oxygen atom (\Oi) to form the
carbon-oxygen interstitial pair (\CiOi)
complex~\cite{1989--Davies--PR--OPLC}, also known as C-center. It
emits light at telecom wavelengths of \qty{1570}{\nm}
(\qty{0.790}{\eV})~\cite{1968--Spry--PR--RLIS,1984--Wagner--PRB--ESCL,1985--Thonke--JPCSSP--CLDI}
and is thus a suitable candidate for a telecom-range single-photon
source compatible with fiber-optic technology, although experimental
confirmation of its SPE capabilities remains absent.

\begin{figure*}
  \centering
  \includegraphics[width=0.6\linewidth]{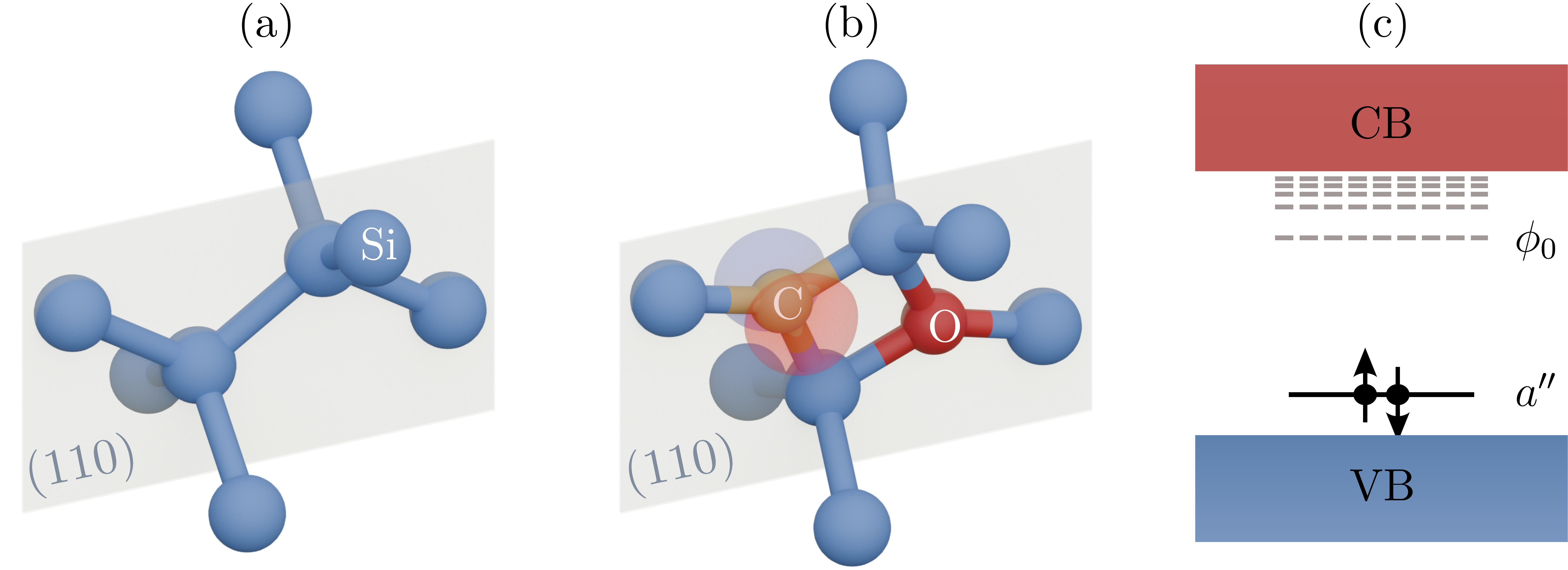}
  \caption{\label{fig:structure} Atomic structures of (a) bulk silicon
    and (b) the \CiOi\ defect, known as the C-center. The mirror plane
    of the \pgroup{C}{s} point group aligns with the (110)
    crystallographic plane and intersects the square-like ring formed
    by two silicon (blue), one interstitial carbon (brown), and one
    interstitial oxygen (red) atom. Picture~(c) shows a qualitative
    Kohn--Sham electronic energy level diagram for the C-center's
    ground state. The $sp^2$ hybridization of the carbon and oxygen
    atoms’ orbitals results in an in-gap defect level, labeled $a''$,
    above the valence band (VB) as a $p$-type orbital localized on the
    carbon atom. Bound exciton states are depicted by dashed gray
    lines below the conduction band (CB), with $\phi_0$ denoting the
    first state in the series.}
\end{figure*}

The geometric configurations of the C-center defect is identified as a
square-like ring within the (110) crystallographic plane, formed by
two silicon atoms and two
interstitials~\cite{1987--Trombetta--MP--IICI,1992--Jones--PRL--OFIC}.
Figures~\ref{fig:structure}(a) and \ref{fig:structure}(b) illustrate
the formation of this defect: Fig.~\ref{fig:structure}(a) shows the
fragment of two silicon atoms along with their nearest neighbors,
while Fig.~\ref{fig:structure}(b) presents the same fragment with the
defect structure in place. This defect exhibits \pgroup{C}{s} point
group symmetry~\cite{1985--Thonke--JPCSSP--CLDI}.

The electronic structure of the neutral C-center defect can be
elucidated from the perspective of the molecular orbitals. As depicted
in Fig.~\ref{fig:structure}(b), both carbon and oxygen atoms form
three $sp^2$ bonds with the nearest silicon atoms, leaving four
electrons to occupy two $p$-type orbitals. For neutral centers, this
configuration yields a closed-shell singlet ground state. Previous
theoretical calculations predict that the only in-gap defect level
$a''$ (where $a''$ denotes the irreducible representation of the
\pgroup{C}{s} point group) is a $p$-type orbital localized on the
carbon atom~\cite{2022--Udvarhelyi--npjCM--LBEQ}, while the
oxygen-related $p$-type orbital is resonant with bulk states. Although
such an electronic structure suggests the absence of an intradefect
transition, photoluminescence excitation measurements reveal a series
of bound exciton
states~\cite{1984--Wagner--PRB--ESCL,1985--Thonke--JPCSSP--CLDI}.
These states arise from the interaction between a charged defect and a
bound carrier, resulting in a hydrogenic wave function of the state.
This, in turn, defines the single-particle energy level structure of
the ground state, as illustrated in Fig.~\ref{fig:structure}(c), where
$a''$ is the defect-induced in-gap state and $\phi_0$ denotes the
first in the series of bound exciton states.

In this work we use first-principles calculations to study the
thermodynamic properties, vibrational structure, and optical activity
of the \CiOi\ defect. This involves the calculation of thermodynamic
parameters, including the formation energies of the most stable
geometries of \Ci, \Oi, and \CiOi\ in different charge states, the
binding energy of a neutral \CiOi\ complex, and charge-state
transition levels (CTL) of \Ci, \Oi, and \CiOi\ defects. We calculate
the luminescence and absorption lineshapes of the neutral C-center in
silicon. To mitigate the impact of finite-sized supercells, we employ
an embedding
methodology~\cite{2014--Alkauskas--NJP--FPTL,2021--Razinkovas--PRB--VVSI},
which allows us to simulate the vibrational structure and the
adiabatic potential energy surfaces of defects in the dilute limit and
model the electron--phonon coupling during an optical transition to a
high degree of accuracy. The presence of resonances, bulk-like, and
localized vibrational modes in our calculations explains various
features of the low-temperature experimental luminescence spectrum.
Furthermore, our results show that high-accuracy lineshape modeling
for exciton-like excitations can be achieved using a simple
approximate methodology where the excited state geometry of a neutral
system is obtained by relaxing the structure of a positively charged
system. Our work provides additional evidence for the correct
assignment of the C-center to the carbon-oxygen interstitial pair
defect. Moreover, it reinforces the idea that optical lineshape
modeling can be a powerful aid in semiconductor point defect
identification.


\section{Methods\label{sec:methods}}

\subsection{First-principles calculations}

Computations were performed using spin-polarized density functional
theory (DFT), as implemented in the Vienna \emph{Ab initio} Simulation
Package (VASP)~\cite{1996--Kresse--PRB--EISA}. The projector-augmented
wave (PAW) method was applied~\cite{1994--Blochl--PRB--PAWM}, with a
plane-wave energy cutoff of \qty{500}{\eV}. Brillouin zone sampling
was restricted to the $\Gamma$ point. For defect modeling, we used a
cubic supercell containing \num{512} atomic sites
($4 \times 4 \times 4$ conventional cells). Phonon calculations were
carried out using the meta-GGA SCAN
functional~\cite{2015--Sun--PRL--SCAN} to account for the exchange and
correlation, while the defect formation energy calculations employed
the HSE06 hybrid functional~\cite{2003--Heyd--JCP--HFBS}. For HSE06, a
default fraction ($a = 0.25$) of screened Fock exchange is combined
with the semilocal exchange-correlation functional of Perdew, Burke,
and Ernzerhof (PBE)~\cite{1996--Perdew--PRL--GGAM}. The above choice
of functionals was motivated by their strengths: SCAN offers an
optimal balance between accuracy and computational efficiency, making
it ideal for capturing the structural and vibrational properties of
deep-level
defects~\cite{2023--Maciaszek--JCP--ASDF,2024--Silkinis--PRB--OLOS}.
HSE06, with its superior accuracy in describing the band gap, was
employed for defect formation energies to accurately determine
thermodynamic properties.

\subsection{Thermodynamic properties}

The formation energy of a defect $X$ in charge state $q$ is calculated
using the following formula~\cite{2014--Freysoldt--RMP--FPCP}:
\begin{equation}
  \label{eq:formation_energy}
  \begin{split}
    \Delta \func{H_{\mathrm{f}}^{X}}{q}
    = {} & \func{E_{\mathrm{tot}}^{X}}{q} - E_{\mathrm{bulk}} - \sum_{i}{n_i \mu_i} \\
         & {} + q (E_{\mathrm{VBM}} + E_{\mathrm{F}}) + E_{\mathrm{corr}}.
  \end{split}
\end{equation}
Here, $\func{E_{\mathrm{tot}}^{X}}{q}$ is the total energy of the
supercell containing the defect $X$, and $E_{\mathrm{bulk}}$ is the
total energy of a defect-free supercell of the same size. The chemical
potential $\mu_i$ corresponds to type $i$ atoms that have been added
to ($n_i > 0$) or removed from ($n_i < 0$) the supercell. The Fermi
energy $E_{\mathrm{F}}$ describes the chemical potential of the
electron and is referenced to the valence band maximum (VBM)
$E_{\mathrm{VBM}}$. Finally, $E_{\mathrm{corr}}$ is a correction term
for the electrostatic interactions between mirror images of charged
defects due to periodic boundary conditions. In this work, we used the
electrostatic correction scheme described in
Ref.~\cite{2009--Freysoldt--PRL--FAFS}.

For defect complexes, an important quantity to consider is their
binding energy. For a complex $AB$, composed of two defects $A$ and
$B$, the binding energy is given by
\begin{equation}
  \label{eq:binding_energy}
  E_{\mathrm{b}}
  = \Delta \func{H_{\mathrm{f}}^{A}}{q} + \Delta \func{H_{\mathrm{f}}^{B}}{q}
  - \Delta \func{H_{\mathrm{f}}^{AB}}{q}.
\end{equation}
A positive binding energy $E_{\mathrm{b}}$ entails that the formation
of a complex is more energetically favorable than having its
constituent parts exist individually. However, it should be noted that
a positive $E_{\mathrm{b}}$ value does not indicate that considerable
concentrations of complexes will be formed because defect complexes
generally have much smaller configurational entropies than their
constituent parts~\cite{2004--Van-de-Walle--JAP--FPCD}.

Defects in solids can exist in different charge states, depending on
the occupation of defect levels within the band gap. To identify the
Fermi level ranges where a specific charge state is stable, we
calculate the charge state transition level
(CTL)~\cite{2014--Freysoldt--RMP--FPCP} as
\begin{equation}
  \label{eq:ctl}
  \varepsilon(q/q')
  = \frac{\Delta \func{H_{\mathrm{f}}^{X}}{q; E_{\mathrm{F}} = 0}
    - \Delta \func{H_{\mathrm{f}}^{X}}{q'; E_{\mathrm{F}} = 0}}
  {q' - q},
\end{equation}
where $\Delta \func{H_{\mathrm{f}}^{X}}{q; E_{\mathrm{F}} = 0}$
denotes the formation energy of defect $X$ in the charge state $q$
with the Fermi level set to zero. The charge state $q$ is stable for
Fermi levels below $\varepsilon(q/q')$, while for Fermi levels above
$\varepsilon(q/q')$, the system stabilizes in the $q'$ charge
state~\cite{2014--Freysoldt--RMP--FPCP}. Additionally, the ionization
energy is related to the CTL, where the energy required to remove an
electron from a defect in charge state $q'$ and transfer it to the
conduction band corresponds to the distance between the
$\varepsilon(q/q')$ and the conduction band minimum (CBM). Conversely,
the energy needed for the defect to capture an electron from the
valence band while in the charge state $q$ is represented by the
distance between $\varepsilon(q/q')$ and the VBM.

\subsection{Electronic structure}

Since the neutrally charged C-center defect is a closed-shell system
[see Fig.~\ref{fig:structure}(c)], its ground state is represented by
a single determinant wave function,
$\ket{S_0} = \ket{a'' \sdown{a}''}$, where both electrons occupy the
only molecular orbital located within the band-gap. In this notation,
the presence (or absence) of a bar symbol denotes the spin-down (or
spin-up) configuration of the molecular orbital. The electron
excitation from the $a''$ orbital to the bound exciton level $\phi_0$
leads to either the excited triplet or singlet state. However,
reaching the triplet state requires a spin flip, making this
transition dipole forbidden. The wave function of the excited triplet
state in the $m_s = 1$ projection is expressed as a single
determinant, $\ket{T} = \ket{a'' \phi_0}$. In contrast, the excited
singlet state is strongly correlated and expressed as a
multi-determinant wave function,
$\ket{S_1} = \frac{1}{\sqrt{2}} (\ket{a'' \sdown{\phi_0}} -
\ket{\sdown{a}'' \phi_0})$. The transition from the ground singlet to
this excited singlet is dipole-allowed; however, this state is
inaccessible via \enquote{traditional} DFT methods due to the
single-determinant nature of Kohn–Sham wave functions.

The exciton-like nature of excited singlet states $\ket{S_{1}}$ adds
additional complexity. Modeling these states using accessible
supercell sizes ($\leqslant \num{1000}$~atoms) is a challenging
problem because the effective radii of such single-particle states are
larger than the dimensions of the supercell. Recently, a semi-\emph{ab
  initio} method for modeling bound exciton states of deep defects was
developed~\cite{2024--Chen--PRB--SAMB}, where the authors of the paper
use an effective mass model with potentials from regular DFT
calculations to capture the structure of these exciton states and
apply it to the negatively charged nitrogen-vacancy center in diamond.
However, this approach does not allow for investigating the adiabatic
potential energy surfaces (APES) required to model spectral
lineshapes.

Describing the potential energy surfaces in both ground and excited
electronic states is sufficient for modeling electron--phonon
interactions. Based on the hypothesis that the form of the APES of a
defect primarily depends on the local charge distribution of localized
orbitals, we propose a simple and practical approach. In this method,
we model the structural properties of the bound exciton state
$\ket{S_{1}}$ by considering the ground state of the positively
charged defect and neglecting the contributions of exciton-like
orbitals. The validity of this approach is confirmed \emph{a
  posteriori} through comparison with experimental results. This
methodology is further supported by a recent study on the \Ci\ defect
in silicon, which similarly approximated the bound exciton state using
a positively charged defect~\cite{2024--Deak--CP--QBTW}. Thus, unless
specified otherwise, it is assumed that the geometry of the
exciton-like excited singlet state $\ket{S_{1}}$ is approximated by
the geometry of the positively charged defect in the ground state.

To further validate our above approach, we employed the \dSCF\
methodology~\cite{2009--Gali--PRL--TSEC} to approximate the geometry
of the excited singlet state by promoting an electron from the $a''$
orbital in the spin-down channel to the $\phi_{0}$ orbital in the
spin-up channel. This excitation nominally forms a triplet state.
However, given that the triplet and singlet excited states share the
same orbital configuration and differ only by a small exchange
splitting due to the electron delocalization, which is about
\qty{3}{\milli\eV}~\cite{1993--Bohnert--PRB--TCIB,2011--Ishikawa--PRB--OPTS},
the geometry of the excited triplet state $\ket{T}$ should closely
mirror that of the excited singlet state $\ket{S_{1}}$.

Both approaches represent limiting cases: the positively charged
geometry neglects the contribution of the bound electron, while the
triplet geometry overestimates the charge density due to the limited
size of the supercell.

\subsection{Vibrational structure and characterization of phonons}

The key parameters required to compute the vibrational structure of a
defect are the elements of the Hessian matrix (also known as the force
constant matrix)
\begin{equation}
  \label{eq:Hessian}
  \Phi_{\alpha,\beta}(m,n)
  = \frac{\partial^2 V}{\partial u_{m\alpha} \partial u_{n\beta}}
  = \frac{\partial F_{m\alpha}}{\partial u_{n\beta}},
\end{equation}
where $V$ represents the potential energy surface for ions,
$u_{m\alpha}$ is the displacement of atom $m$ along direction
$\alpha$, and $F_{m\alpha}$ is the force component on atom $m$ along
direction $\alpha$. In this study, the Hessian matrix was computed
using the finite-difference method~\cite{1995--Kresse--EL--AFCA},
which involves calculating forces $F_{m\alpha}$ induced by small
atomic displacements $u_{n\beta}$, systematically displacing each atom
in the supercell. This approach requires a large number of
single-point calculations for a defect-containing supercell with
broken translational symmetry. However, the computational effort was
reduced by exploiting the point-group symmetry of the system, as
implemented in the \texttt{phonopy} software
package~\cite{2023--Togo--JPSJ--FPPC,2023--Togo--JPCM--ISPP}. Once the
Hessian matrix is obtained, diagonalization of the dynamical matrix
$D_{\alpha,\beta}(m,n) = \Phi_{\alpha,\beta}(m,n)/\sqrt{M_{m}M_{n}}$,
where $M_{m}$ and $M_{n}$ are the atomic masses, yields the
mass-weighted normal modes $\vec{\eta}_{k}$ and vibrational
frequencies $\omega_{k}$ of the defect.

The finite-size effects inherent in supercells with
\mbox{$\leqslant \num{1000}$}~atoms hinder accurate modeling of the
vibrational properties of point defect systems. The use of periodic
boundary conditions in these relatively small supercells artificially
increases the defect concentration and restricts the representation of
long-wavelength acoustic phonons. To address these limitations, we
employ an advanced embedding methodology that constructs large Hessian
matrices from smaller, directly accessible
supercells~\cite{2014--Alkauskas--NJP--FPTL,2021--Razinkovas--PRB--VVSI}.
This approach leverages the short-range nature of interatomic forces,
enabling the calculation of vibrational properties within much larger
effective supercells by combining Hessian matrix elements from both
defect-containing and bulk supercells. For a more detailed description
of this methodology, we refer to Appendix~\ref{app:embedding}, as well
as the original works in
Refs.~\cite{2014--Alkauskas--NJP--FPTL,2021--Razinkovas--PRB--VVSI}.

Further understanding of the defect's vibrational properties can be
obtained by analyzing the \enquote{localization} of the phonon
modes~\cite{1970--Bell--JPCSSP--LNMV}. Bulk-like modes are spatially
delocalized, resembling plane wave-like vibrations, while localized
modes correspond to vibrations confined near the defect site, with
frequencies outside the bulk phonon band. Between these two extremes
are quasi-local modes~\cite{2014--Alkauskas--NJP--FPTL}, which are
defect-induced vibrational resonances. These modes arise from a
collection of vibrations whose energy values fall within the range of
bulk phonons in the pristine material.

To quantify the phonon mode localization, we use the concept of an
\enquote{inverse participation ratio} (IPR), which is defined for each
phonon mode $k$
as~\cite{1970--Bell--JPCSSP--LNMV,2014--Alkauskas--NJP--FPTL}
\begin{equation}
  \label{eq:ipr}
  \mathrm{IPR}_{k} = \frac{1}{\sum_{m}\vec{\eta}_{k;m}^{4}},
\end{equation}
where $\vec{\eta}_{k;m}$ is the three-dimensional mass-weighted
displacement of atom $m$, and
$\vec{\eta}_{k;m}^{4} \equiv
(\sum_{\alpha}{\eta_{k;m\alpha}^{2}})^{2}$, with
$\eta_{k;m\alpha}$ representing the normalized mass-weighted
displacement of atom $m$ along direction $\alpha$ for phonon mode $k$.
The value of $\mathrm{IPR}_{k}$ measures the number of oscillating
atoms in vibrational mode $\vec{\eta}_{k}$. For example,
$\mathrm{IPR} = 1$ if only a single atom vibrates for a given mode.
Conversely, $\mathrm{IPR} = N$ if all $N$ atoms in the supercell
participate equally in the vibration. For intermediate cases, where
only $P < N$ atoms vibrate appreciably, $\mathrm{IPR} \approx P$. A
constant $\mathrm{IPR}$ as a function of the supercell size $N$
characterizes a truly localized mode. This behavior occurs because the
frequency of the localized mode lies outside the bulk phonon band, and
the mode's spatial profile remains independent of the density or
presence of bulk phonons.

While IPR provides a quantitative measure of phonon mode localization,
it is more convenient to work with the \enquote{localization ratio}
$\beta_{k}$ of phonon mode $k$~\cite{2014--Alkauskas--NJP--FPTL}:
\begin{equation}
  \label{eq:beta}
  \beta_{k} = \frac{N}{\mathrm{IPR}_{k}}.
\end{equation}
As seen from this definition, the larger the ratio $\beta_{k}$, the
more localized the phonon mode is.

\subsection{Optical lineshapes and electron--phonon coupling parameters\label{subsec:lineshapes}}

In the Franck--Condon and adiabatic approximations, the normalized
lineshape $\func{L}{\hbar\omega}$ for the absorption and emission
spectra of a semiconductor defect as a function of frequency at
$T = \qty{0}{\kelvin}$ is given by~\cite{2021--Razinkovas--PRB--VVSI}
\begin{align}
  \label{eq:lineshape}
  \func{L}{\hbar\omega}
  & \! = C \omega^{\kappa} \func{A}{\hbar\omega}, \\
  \label{eq:spectral_function}
  \func{A}{\hbar\omega}
  & \! = \sum_{m} \! {
    \abs{\braket{\chi_{i0}}{\chi_{fm}}}^2
    \func{\delta}{
    E_{\mathrm{ZPL}}
    {\mp} (\varepsilon_{fm} \! - \! \varepsilon_{f0})
    \! - \! \hbar\omega}
    },
\end{align}
where $C$ is a normalization constant, $E_{\mathrm{ZPL}}$ is the
zero-phonon line (ZPL) energy, and $\chi_{i0}$ and $\chi_{fm}$ are the
vibrational wave functions of the initial ($i$) and final ($f$)
electronic states. Here, $\varepsilon_{fm}$ is the energy of the
$m$-th vibrational state in the final electronic manifold with respect
to the potential energy minima. The minus and plus signs in the
argument of the $\delta$-function correspond to emission and
absorption, respectively, with the parameter $\kappa$ taking the value
of $3$ for emission and $1$ for absorption. The optical spectral
function $\func{A}{\hbar\omega}$ quantifies the transition amplitudes
between vibrational states and is central in determining the
lineshape.

To simplify the calculation of the overlap integrals
$\braket{\chi_{i0}}{\chi_{fm}}$ in Eq.~\eqref{eq:spectral_function},
we adopt the equal-mode
approximation~\cite{2021--Razinkovas--PRB--VVSI,1959--Markham--RMP--INME},
which assumes that the vibrational modes of the initial and final
states are identical. To address the limitations of this assumption,
we follow Ref.~\cite{2021--Razinkovas--PRB--VVSI} and consistently
employ the vibrational modes of the final state, using the ground
state phonons for emission and the excited state phonons for
absorption.

Efficient estimation of the optical spectral function
$\func{A}{\hbar\omega}$ can be obtained by computing it in the time
domain through the generating function method of Kubo and
Lax~\cite{1955--Kubo--PTP--AMGF,1952--Lax--JCP--FCPI}. In this
approach, the spectral function is obtained from the generating
function $\func{G}{t}$ via the relation
\begin{equation}
  \label{eq:opt_spectral_func_gen}
  \func{A}{\hbar\omega}
  = \frac{1}{2 \pi} \int_{-\infty}^{\infty}{
    \func{G}{t} \en^{-\gamma \abs{t}} \en^{- \iu (E_{\mathrm{ZPL}} / \hbar - \omega) t} \odif{t}
  }.
\end{equation}
The term $\en^{-\gamma \abs{t}}$ is introduced as a
phenomenological correction to account for the homogeneous Lorentzian
broadening of the ZPL not captured by the theory, as well as to
address inhomogeneous broadening. In practice, parameter $\gamma$ is
adjusted to match the experimental linewidth of the ZPL.

In the equal mode approximation, the generating function can be
expressed as
\begin{equation}
  \label{eq:gen_func}
  \func{G}{t}
  = \exp \left[ -\sum_{k}{S_{k}} + \sum_{k}{S_{k}e^{\pm \iu \omega_{k} t}} \right],
\end{equation}
where the plus sign in the exponent is used for emission and the minus
sign for absorption, $S_k$ represents the partial Huang--Rhys (HR)
factor, and the summation runs over all vibrational modes of the
system. The factor $S_k$ quantifies the average number of excited
phonons during an optical transition~\cite{1950--Huang--PRSLA--TLAN}.
It is defined as
\begin{equation}
  \label{eq:part_hr_factor}
  S_k = \frac{\omega_k \Delta Q_{k}^{2}}{2 \hbar},
\end{equation}
where $\Delta Q_{k}$ represents the ionic displacement along the
normal mode $k$ induced by an optical transition. Specifically,
$\Delta Q_{k}$ is the projection of the mass-weighted displacement
between the ground and excited states onto the normalized phonon mode
$\vec{\eta}_{k}$:
\begin{equation}
  \label{eq:delta_q}
  \Delta Q_{k} = \sum_{m\alpha}{\sqrt{M_{m}} \Delta R_{m\alpha} \eta_{k; m\alpha}}.
\end{equation}
Here, $\Delta R_{m\alpha}$ is the displacement of the atom $m$ along
the $\alpha$ coordinate, and $M_{m}$ is its mass.

A primary challenge in evaluating Eq.~\eqref{eq:gen_func} for extended
systems, such as defects, is that the electron--phonon coupling must
account for a continuum of vibrational frequencies. To address this
and achieve a converged, continuous description of the
electron--phonon interaction, we first define the spectral density of
the electron--phonon coupling (also known as the spectral function of
electron--phonon coupling) as
\begin{equation}
  \label{eq:3}
  \func{S}{\hbar\omega}
  \equiv \sum_{k} S_{k} \func{\delta}{\hbar\omega_{k} - \hbar\omega},
\end{equation}
which allows us to rewrite the generating function as follows:
\begin{equation}
  \label{eq:4}
  \func{G}{t}
  = \exp \left[ -S_{\mathrm{tot}} + \int \func{S}{\hbar\omega} e^{\pm \iu \omega t} \odif{\omega} \right].
\end{equation}
Here, $S_{\mathrm{tot}} = \sum_{k} S_{k}$ is the total HR factor,
where the sum is evaluated over all vibrational modes of the defect
system. We employ smoothing functions tailored to different mode types
to approximate the $\delta$-functions in Eq.~\eqref{eq:3}. For
bulk-like and quasi-localized modes, $\delta$-functions are replaced
with Gaussians of specified width, which allows controlled spectral
broadening. For highly localized modes, we instead use Lorentzians,
with half-width at full maximum (HWHM) chosen to accurately capture
the side peaks in the optical lineshape. For a given spectral
resolution, we apply the embedding methodology and check the
convergence of the optical lineshape as a function of the supercell
size (see Fig.~\ref{fig:luminescence_embedded_convergence} in
Appendix~\ref{app:embedding} for an illustration). Usually, the
convergence is achieved with a system size greater than
\num{10000}~atoms.

An additional complication in supercell calculations arises when
determining the relaxation profile following an optical transition.
The small supercell size in direct calculations suppresses the
long-wavelength components of $\Delta R_{m\alpha}$ in
Eq.~\eqref{eq:delta_q}. To accurately capture the relaxation profile
in the dilute limit, we calculate $\Delta Q_{k}$ for each vibrational
mode of a large supercell (obtained via the embedding procedure) by
using forces and the harmonic relation to displacements. The
relaxation component $\Delta Q_{k}$ is then evaluated as
\begin{equation}
  \label{eq:5}
  \Delta Q_{k}
  = \frac{1}{\omega_{k}^{2}} \sum_{m\alpha}\frac{F_{m\alpha}}{\sqrt{M_{m}}} \eta_{k; m\alpha},
\end{equation}
where $F_{m\alpha}$ represents the force on atom $m$ along direction
$\alpha$ when the system is in the final electronic state but remains
in the equilibrium geometry of the initial state calculated in the
directly accessible supercell. Meanwhile, $\eta_{k; m\alpha}$
represents the vibrational mode shape as obtained in the embedded
supercell. With the forces already converged within the explicitly
accessible supercell, this approach, combined with the embedding
methodology, effectively captures the electron--phonon coupling for
low-frequency modes and provides an accurate description of the
optical lineshapes.


\section{Results and discussion\label{sec:results}}

\subsection{Lattice parameters and band gaps}

We begin by calculating the lattice constants $a$ and band gaps
$E_{\mathrm{g}}$ of pure silicon using the geometry relaxation
procedures with PBE, SCAN, and HSE06 functionals. Table~\ref{tab:bulk}
compares these results with low-temperature experimental
values~\cite{1964--Batchelder--JCP--LCTE,1974--Bludau--JAP--TDBG}.

\begin{table}
  \renewcommand{\arraystretch}{0.7}
  \caption{\label{tab:bulk} Calculated lattice constants $a$ (in
    \unit{\angstrom}) and band gaps $E_{\mathrm{g}}$ (in \unit{\eV})
    of pure silicon with different functionals compared to
    low-temperature experimental data.}
  \begin{tabular}{l @{\hspace{2em}} S[table-format=1.3] @{\hspace{2em}} S[table-format=1.3]}
    \toprule
    & {$a$} & {$E_{\mathrm{g}}$} \\
    \midrule
    PBE   & 5.469 & 0.611 \\[2ex]
    SCAN  & 5.428 & 0.825 \\[2ex]
    HSE06 & 5.433 & 1.153 \\[2ex]
    Expt. & 5.419\NoteMark{a} & 1.170\NoteMark{b} \\
    \bottomrule
  \end{tabular}\\[1ex]
  {\footnotesize
    \NoteMark{a}Ref.~\cite{1964--Batchelder--JCP--LCTE}\hspace{2em}
    \NoteMark{b}Ref.~\cite{1974--Bludau--JAP--TDBG}
  }
\end{table}

The SCAN functional yields the best agreement with the experimental
lattice constant, providing the value of $a = \qty{5.428}{\angstrom}$
with an absolute deviation of \qty{0.009}{\angstrom}. In comparison,
the HSE06 functional produces the lattice constant of
$a = \qty{5.433}{\angstrom}$ with an absolute deviation of
\qty{0.014}{\angstrom}, while the PBE functional overestimates this
parameter, yielding $a = \qty{5.469}{\angstrom}$ with a deviation of
\qty{0.050}{\angstrom}. Regarding the band gaps, the HSE06 functional
shows the best agreement with the experiment, with an absolute
deviation of \qty{0.017}{\eV} and a value of
$E_{\mathrm{g}} = \qty{1.153}{\eV}$. PBE and SCAN underestimate the
band gap by \qty{0.559}{\eV} and \qty{0.345}{\eV}, respectively, a
common feature of semilocal functionals. Based on the above results,
the SCAN functional was selected for defect phonon and relaxation
profile calculations due to its accuracy in predicting structural
parameters and computational
efficiency~\cite{2023--Maciaszek--JCP--ASDF}. The more computationally
demanding HSE06 functional was used exclusively for formation energy
calculations.

\subsection{Thermodynamic properties}

The calculated formation energies of neutral \Ci, \Oi, and \CiOi\
defects are summarized in Table~\ref{tab:formation_energies},
alongside comparisons to previous studies. These values were
calculated using Eq.~\eqref{eq:formation_energy}, with the chemical
potentials for Si, C, and O obtained from silicon, diamond, and
$\alpha$-quartz (SiO\textsubscript{2}), respectively. To our
knowledge, experimental reference data for defect formation energies
are limited. However, a key comparison is available for the \Oi\
defect from~Ref.~\cite{1971--Bean--JPCS--SCPS}.

\begin{table}
  \caption{\label{tab:formation_energies} Calculated formation
    energies $\Delta H_{\mathrm{f}}^{X}(0)$ (in~\unit{\eV}) of neutral
    \Ci, \Oi, and \CiOi\ defects in silicon compared to previous
    theoretical and experimental results. The only available
    experimental reference for defect formation energy is for the \Oi\
    defect.}
  \begin{tabular}{l @{\hspace{1em}} S[table-format=1.2] @{\hspace{1em}} S[table-format=1.2] @{\hspace{1em}} S[table-format=1.2]}
    \toprule
    & \multicolumn{1}{c}{$\Delta H_{\mathrm{f}}^{\Ci}(0)$}
    & \multicolumn{1}{c}{$\Delta H_{\mathrm{f}}^{\Oi}(0)$}
    & \multicolumn{1}{c}{$\Delta H_{\mathrm{f}}^{\CiOi}(0)$} \\
    \midrule
    SCAN (this work)  & 3.49 & 1.79 & 3.75 \\[3ex]
    HSE06 (this work) & 3.70 & 1.81 & 3.95 \\[3ex]
    \multirow{4}{*}{Prior theory} & 4.12\NoteMark{a} & 1.81\NoteMark{a} & 4.21\NoteMark{a} \\
    & 4.50\NoteMark{b} & 1.95\NoteMark{b} & 4.85\NoteMark{b} \\
    & 3.74\NoteMark{c} & 1.56\NoteMark{d} & 3.63\NoteMark{c} \\
    & 3.72\NoteMark{e,f} & {--} & {--} \\[3ex]
    Expt. & {--} & \num{1.65(0.15)}\NoteMark{g} & {--} \\
    \bottomrule
  \end{tabular}\\[1ex]
  {\footnotesize
    \NoteMark{a}Ref.~\cite{2001--Coutinho--PRB--ICOC}\quad
    \NoteMark{b}Ref.~\cite{2014--Wang--SR--CRDI}\quad
    \NoteMark{c}Ref.~\cite{2004--Hao--JPCM--ICDB}\quad
    \NoteMark{d}Ref.~\cite{2004--Hao--PRB--IOSS}\\
    \NoteMark{e}Ref.~\cite{2011--Zirkelbach--PRB--CACP}\quad
    \NoteMark{f}Ref.~\cite{2024--Deak--CP--QBTW}\quad
    \NoteMark{g}Ref.~\cite{1971--Bean--JPCS--SCPS}
  }
\end{table}

In the case of the neutral \Ci\ defect, we considered three distinct
geometrical configurations, following
Ref.~\cite{2011--Zirkelbach--PRB--CACP}: the $\abra{100}$ \Ci\
dumbbell, the $\abra{110}$ \Ci\ dumbbell, and the bond-centered~\Ci.
In agreement with the results from
Refs.~\cite{1976--Watkins--PRL--EOII,1990--Song--PRB--EISS,2011--Zirkelbach--PRB--CACP,2024--Deak--CP--QBTW},
our calculations indicate that the $\abra{100}$ dumbbell configuration
is the most energetically favorable. The formation energy values of
\qty{3.49}{\eV} and \qty{3.70}{\eV} were obtained using the SCAN and
HSE06 functionals, respectively. The calculated formation energies of
the $\abra{110}$ and bond-centered configurations are larger by
\qty{0.55}{\eV} (\qty{0.49}{\eV}) and \qty{0.70}{\eV}
(\qty{0.81}{\eV}) using the SCAN (HSE06) functional. Unless otherwise
specified, the \Ci\ in the remainder of this text refers to the
$\abra{100}$ \Ci\ dumbbell configuration.

The calculated formation energy values of the neutral \Oi\ defect
become \qty{1.79}{\eV} and \qty{1.81}{\eV} for the SCAN and HSE06
functionals, respectively. These values are consistent with the upper
limit of the \qty{1.65(15)}{\eV} experimental value from
Ref.~\cite{1971--Bean--JPCS--SCPS}, thereby validating the accuracy of
our computational methodologies.

By applying Eq.~\ref{eq:binding_energy} and utilizing the data from
Table~\ref{tab:formation_energies}, the calculated binding energies of
the neutral \CiOi\ complex are \qty{1.53}{\eV} and \qty{1.56}{\eV} for
the SCAN and HSE06 functionals, respectively. These values show good
agreement with previous theoretical
studies~\cite{2001--Coutinho--PRB--ICOC,2004--Hao--JPCM--ICDB,2008--Backlund--PRB--CDPS},
which report binding energies in the range of \qty{1.6}{\eV} to
\qty{1.7}{\eV}.

\begin{table}
  \caption{\label{tab:ctls} Calculated charge-state transition levels
    (CTLs, in~\unit{\eV}) of the \Ci\ and \CiOi\ defects in silicon
    compared to previous results. HSE06 calculations reveal no CTL
    values for the \Oi\ defect.}
  \begin{tabular}{l @{\hspace{1em}} c @{\hspace{1em}} c}
    \toprule
    & \Ci & \CiOi \\
    \midrule
    \multirow{2}{*}{\makecell[l]{HSE06\\ (this work)}} & $\varepsilon(+/0)\! =\! E_{\mathrm{VBM}}\!+\!0.32$
          & \multirow{2}{*}{$\varepsilon(+/0)\! =\! E_{\mathrm{VBM}}\!+\!0.39$} \\
    & $\varepsilon(0/-)\! =\! E_{\mathrm{CBM}}\!-\! 0.16$ & \\[3ex]

    \multirow{2}{*}{\makecell[l]{Prior\\ theory}}
        & $\varepsilon(+/0)\! =\! E_{\mathrm{VBM}}\!+\!0.32$\NoteMark{a}
        & \multirow{2}{*}{$\varepsilon(+/0)\! =\! E_{\mathrm{VBM}}\!+\!{}$
          \makecell[l]{0.36\NoteMark{b} \\ 0.41\NoteMark{c}}} \\
    & $\varepsilon(0/-)\! =\! E_{\mathrm{CBM}}\!-\! 0.19$\NoteMark{a} & \\[3ex]

    \multirow{2}{*}{Expt.}
        & $\varepsilon(+/0)\! =\! E_{\mathrm{VBM}}\!+\!0.28$\NoteMark{d}
        & \multirow{2}{*}{$\varepsilon(+/0)\! =\! E_{\mathrm{VBM}}\!+\!{}$
          \makecell[l]{0.38\NoteMark{e} \\ 0.36\NoteMark{f}}} \\
    & $\varepsilon(0/-)\! =\! E_{\mathrm{CBM}}\! -\! 0.10$\NoteMark{d} & \\
    \bottomrule
  \end{tabular}\\[1ex]
  {\footnotesize
    \NoteMark{a}Ref.~\cite{2024--Deak--CP--QBTW}\quad
    \NoteMark{b}Ref.~\cite{2001--Coutinho--PRB--ICOC}\quad
    \NoteMark{c}Ref.~\cite{2022--Udvarhelyi--npjCM--LBEQ}\\
    \NoteMark{d}Ref.~\cite{1990--Song--PRB--BICS}\quad
    \NoteMark{e}Ref.~\cite{1977--Mooney--PRB--DELB}\quad
    \NoteMark{f}Ref.~\cite{2019--Ayedh--PSSA--AKIC}
  }
\end{table}

\begin{figure}
  \centering
  \includegraphics{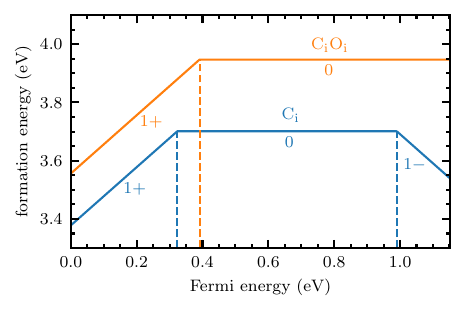}
  \caption{\label{fig:ctls} Formation energy of the \Ci\ and \CiOi\
    defects as a function of the Fermi level for the HSE06
    calculations.}
\end{figure}

Finally, Table~\ref{tab:ctls} and Fig.~\ref{fig:ctls} present the CTLs
of the \Ci\ and \CiOi\ defects in silicon obtained using the HSE06
functional. Our results are compared with previous theoretical studies
from
Refs.~\cite{2001--Coutinho--PRB--ICOC,2022--Udvarhelyi--npjCM--LBEQ,2024--Deak--CP--QBTW},
as well as with available experimental data from
Refs.~\cite{1990--Song--PRB--BICS,1977--Mooney--PRB--DELB,2019--Ayedh--PSSA--AKIC}.
Notably, our calculations indicate no energy levels within the band
gap for the \Oi\ defect; therefore, no CTL data for this defect is
presented.

For the \Ci\ defect, the HSE06 functional yielded a donor
$\varepsilon(+/0)$ CTL at \qty{0.32}{\eV} above the VBM, which aligns
well with the experimental value of
\qty{0.28}{\eV}~\cite{1990--Song--PRB--BICS}. The calculated acceptor
$\varepsilon(0/-)$ transition level is positioned at \qty{0.16}{\eV}
below the CBM, showing good agreement with the experimental value of
\qty{0.10}{\eV}~\cite{1990--Song--PRB--BICS}.

For the \CiOi\ complex, the HSE06 calculations predict a donor
$\varepsilon(+/0)$ CTL at \qty{0.39}{\eV} above the VBM, which agrees
well with experimental values of \qty{0.38}{\eV} and
\qty{0.36}{\eV}~\cite{1977--Mooney--PRB--DELB,2019--Ayedh--PSSA--AKIC}.
Furthermore, this result is consistent with earlier theoretical
predictions of
Refs.~\cite{2001--Coutinho--PRB--ICOC,2022--Udvarhelyi--npjCM--LBEQ},
which place the transition levels at \qty{0.41}{\eV} and
\qty{0.36}{\eV} above the VBM, respectively.

The computed formation energy of the \CiOi\ complex is relatively
high. Assuming that defect concentrations are established at the
silicon melting point (\qty{1687}{\kelvin}), the equilibrium density
of \CiOi\ is expected to remain below \qty{e12}{\per\cubic\cm}.
Furthermore, during the cooling of the sample, if \Ci\ is mobile, it
can bind with \Oi\ to form the \CiOi\ complex due to the relatively
high and positive binding energy of the complex, which further
increases the density of \CiOi\ centers. However, \CiOi\ defects are
typically formed under non-equilibrium conditions, such as during
irradiation, which leads to a higher defect
density~\cite{1996--Zhao--MP--IDRS}.

The ionization threshold energy is given by
$\mathrm{IP} = E_{\mathrm{g}} - \varepsilon(+/0) = \qty{0.78}{\eV}$
(using the experimental value for $E_{\mathrm{g}}$), which represents
the theoretical upper bound for the ZPL energy. According to
theoretical estimates in Ref.~\cite{1985--Thonke--JPCSSP--CLDI}, the
binding energy of the lowest bound exciton level is about
\qty{40}{\milli\eV}, predicting a ZPL energy of \qty{0.74}{\eV}. This
prediction shows good agreement with the experimental ZPL value of
\qty{0.790}{\eV}.

\subsection{Vibrational structures\label{subsec:vibrations}}

\begin{figure}
  \centering
  \includegraphics[width=\linewidth]{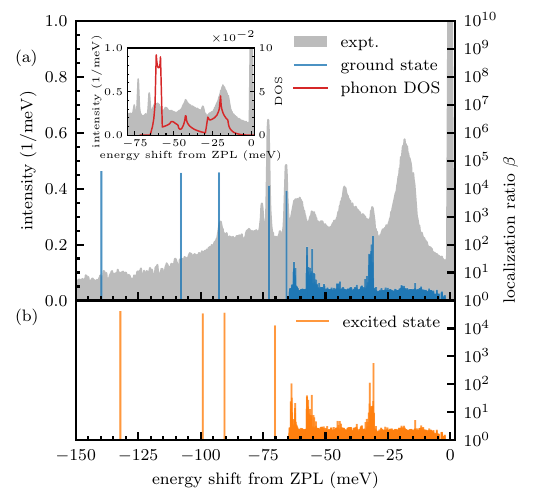}
  \caption{\label{fig:vibrational_structure_embedded} Vibrational
    structures of the (a) ground and (b) excited states of the
    C-center in comparison with normalized experimental luminescence
    data (gray). Energies are taken relative to the ZPL value. Blue
    and orange vertical lines correspond to localization ratios
    $\beta$ of symmetric $a'$ vibrational modes of the
    $18 \times 18 \times 18$ supercell (\num{46656} atomic sites) of
    the ground and excited state, respectively. The red line [inset of
    (a)] corresponds to the density of states (DOS) of phonons in pure
    silicon of the $4 \times 4 \times 4$ supercell (\num{512} atomic
    sites). Experimental data are taken from
    Ref.~\cite{2021--Tajima--APE--FBEI}.}
\end{figure}

Figure~\ref{fig:vibrational_structure_embedded}(a) presents the
calculated localization ratios ($\beta$) for the $a'$ symmetry
vibrational modes (blue) of the ground state alongside normalized
experimental luminescence data (gray) from
Ref.~\cite{2021--Tajima--APE--FBEI}. Higher $\beta$ values indicate
modes with more localized vibration amplitudes near the defect. The
$a''$ vibrational modes are not shown, as they do not contribute to
the linear electron--phonon interaction due to symmetry constraints.
The smaller inset figure overlays the phonon density of states (DOS)
for pristine silicon (red) with the experimental luminescence
spectrum. Note that the horizontal axis in both the main and the inset
figures represents the energy shift relative to the ZPL. The
vibrational modes of the defect system were computed using an
embedding methodology to construct a $18 \times 18 \times 18$
supercell (\num{46656} atomic sites), while the DOS was calculated for
a $4 \times 4 \times 4$ bulk silicon supercell (\num{512} atomic
sites), both employing the SCAN functional. The embedding parameters
and convergence with respect to the supercell size are discussed in
Appendix~\ref{app:embedding}.

\begin{figure}
  \centering
  \includegraphics[width=\linewidth]{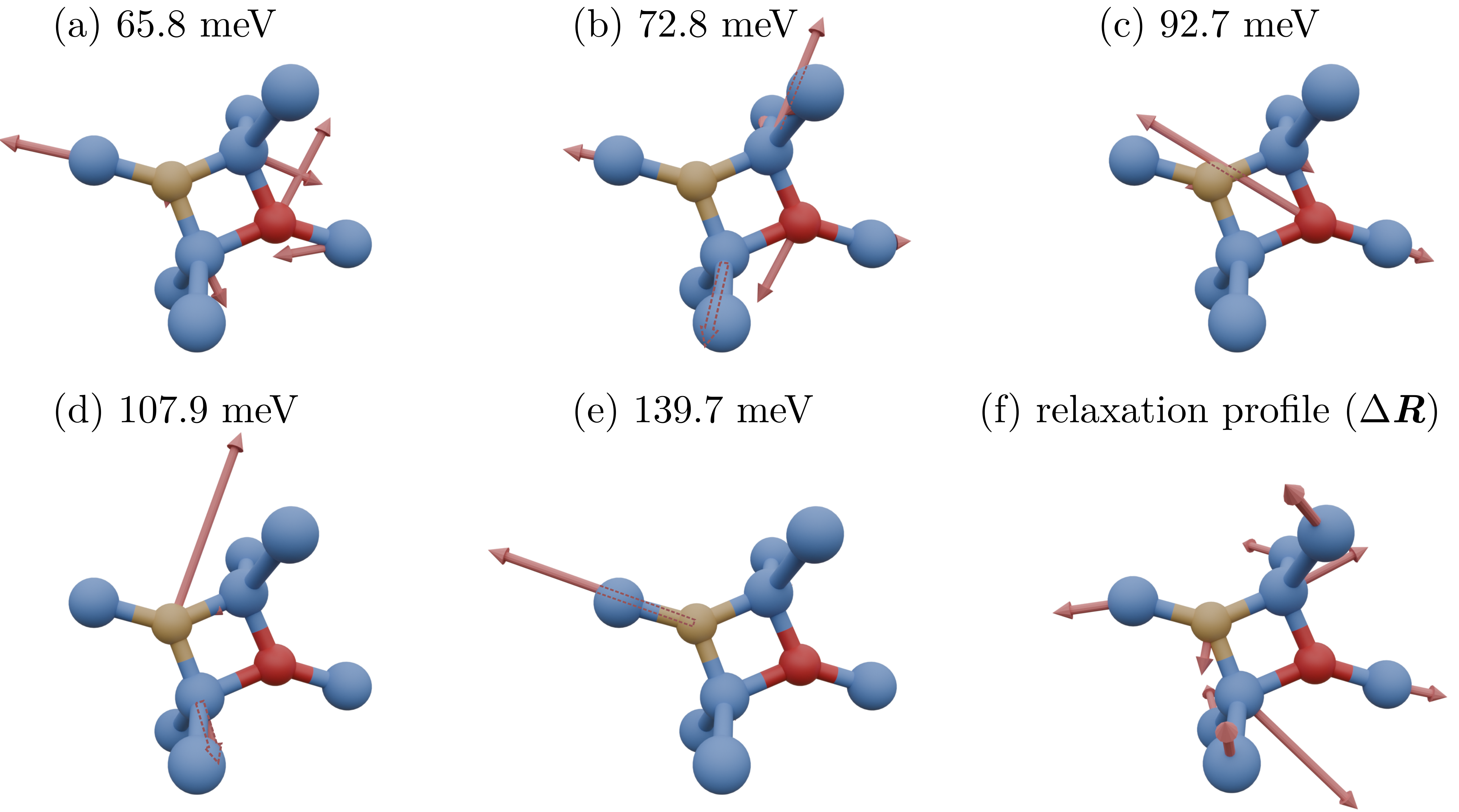}
  \caption{\label{fig:modes} Illustrations of the localized
    vibrational modes of $a'$ symmetry associated with the ground
    state of the C-center defect. Blue spheres correspond to silicon
    atoms, while the brown and red spheres are carbon and oxygen
    interstitials, respectively. The arrows highlight the
    contributions from the nearest neighbor atoms surrounding the ring
    structure. Each vibrational mode's frequency (in \unit{\milli\eV})
    is indicated above its corresponding illustration. Panel~(f)
    depicts the structural changes following the optical excitation.}
\end{figure}

In Fig.~\ref{fig:vibrational_structure_embedded}(a), localized
vibrational modes with high localization ratios
($\beta \geqslant 100$) are identified at energies of
\qtylist{65.8;72.8;92.7;107.9;139.7}{\milli\eV}. Their corresponding
atomic displacements, considering only nearest neighbors, are
illustrated in Figs.~\ref{fig:modes}(a)--(e). Notably, the modes at
\qtylist{65.8;72.8;92.7}{\milli\eV} [Figs.~\ref{fig:modes}(a)--(c)]
show a strong correlation with pronounced peaks in the sideband of the
experimental emission spectrum. In contrast, the lack of luminescence
peaks at \qtylist{107.9;139.7}{\milli\eV} suggests that high $\beta$
values alone do not necessarily correlate with significant
electron--phonon coupling; the mode must also effectively project onto
the relaxation profile $\Delta \vec{R}$ after the optical transition,
as described by Eq.~\eqref{eq:delta_q}. Figure~\ref{fig:modes}(f)
illustrates the calculated equilibrium geometry change following the
excitation, where the positively charged state approximates the
excited singlet state $\ket{S_1}$ of the neutral defect.

The modes visible in the emission spectrum, particularly those at
\qtylist{65.8;72.8;92.7}{\milli\eV}, exhibit considerable vibrational
amplitudes on the oxygen atom and adjacent silicon atoms [see
Figs.~\ref{fig:modes}(a)--(c)]. Although the oxygen atom undergoes
minimal displacement in the relaxation profile shown in
Fig.~\ref{fig:modes}(f), the neighboring silicon atoms experience
notable geometry changes. These displacements align well with the
vibrational amplitudes of the \qtylist{65.8;72.8}{\milli\eV} modes
and, to a lesser extent, the \qty{92.7}{\milli\eV} mode. In contrast,
the higher-frequency modes at \qtylist{107.9;139.7}{\milli\eV} [see
Figs.~\ref{fig:modes}(d) and \ref{fig:modes}(e)] predominantly involve
vibrations centered on the carbon atom, which undergoes only a minor
geometry shift during the optical transition, with minimal mode
participation onto the neighboring silicon atoms. This observation
qualitatively explains the absence of corresponding features for these
higher-frequency modes in the experimental luminescence spectrum.

Beyond the localized vibrations,
Fig.~\ref{fig:vibrational_structure_embedded}(a) reveals other
distinct classes of vibrational modes. The first category, identified
as vibrational resonances, appears as collections of quasi-localized
vibrations centered around \qtylist{32;54;62}{\milli\eV}. These modes
are characterized by $\beta > 10$ values and can be tentatively
associated with corresponding features in the experimental spectrum,
indicating their potential role in coupling with electronic
transitions.

In contrast, bulk-like modes, characterized by relatively small
$\beta$ values ($\beta \leqslant 5$), extend to energies of about
\qty{65}{\milli\eV} from the ZPL, corresponding to the phonon band
edge in pristine silicon. These modes primarily reflect the collective
vibrational characteristics of the silicon lattice and, therefore, are
present in both pure and defect-containing systems. Notably, bulk-like
modes at energies around \qtylist{20;42}{\milli\eV} exhibit peaks in
the phonon DOS function on the low-energy side of the ZPL [see inset
of Fig.~\ref{fig:vibrational_structure_embedded}(a)]. These peaks
coincide with features in the experimental luminescence spectrum,
emphasizing the bulk-like character of these modes and their influence
on the observed spectrum.

In Fig.~\ref{fig:vibrational_structure_embedded}(b), we present the
frequencies and localization factors of vibrational modes calculated
for the positively charged defect, which approximates the excited
state’s structural properties. A comparison between
Fig.~\ref{fig:vibrational_structure_embedded}(a) and
Fig.~\ref{fig:vibrational_structure_embedded}(b) reveals a pronounced
lowering in the vibrational frequencies of localized modes. Most
notably, the mode observed at \qty{65.8}{\milli\eV} in the ground
state shifts to a frequency below the bulk phonon band maximum in the
excited state, transitioning into a vibrational resonance. In
contrast, the quasi-localized vibrations at
\qtylist{32;54;62}{\milli\eV} remain at nearly the same energies in
both ground and excited states. This significant frequency shift of
localized modes indicates a notable quadratic electron--phonon
interaction, which we do not address in this study.

The above analysis shows that the C-center's optical sideband features
can be characterized by examining different vibrational modes.
However, comprehensive calculations of the electron--phonon coupling
during optical processes are necessary to obtain a deeper
understanding of how these modes affect the spectral lineshapes.

\subsection{Optical lineshapes}

In the final part of this study, we analyze the theoretical optical
lineshapes of the C-center in silicon. The accurate calculation of
these lineshapes requires electron--phonon coupling parameters, which
were computed using the same supercell size, exchange-correlation
functional, and vibrational modes as those described in
Section~\ref{subsec:vibrations}.

\begin{figure}
  \centering
  \includegraphics[width=\linewidth]{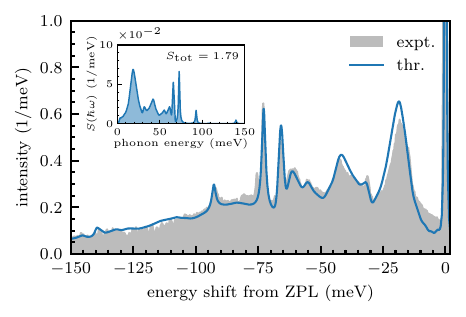}
  \caption{\label{fig:luminescence_embedded} Experimental (gray) and
    calculated (blue) normalized luminescence lineshapes of the
    C-center in silicon. Energies are taken relative to the ZPL value.
    The inset figure contains the calculated spectral function of
    electron--phonon coupling during luminescence. The results
    correspond to calculations for the $18 \times 18 \times 18$
    supercell (\num{46656} atomic sites). Experimental data are taken
    from Ref.~\cite{2021--Tajima--APE--FBEI}.}
\end{figure}

First, by analyzing the calculated partial HR factors [see
Eq.~\eqref{eq:part_hr_factor}] pertaining to localized modes, we
observe significant contributions from the localized vibrational modes
at \qtylist{65.8;72.8;92.7}{\milli\eV}
\mbox{[Figs.~\ref{fig:vibrational_structure_embedded}(a)--(c)]}, with
corresponding HR factors of \numlist{0.126;0.167;0.042}, respectively.
In contrast, vibrational modes involving primarily the carbon atom
exhibit much weaker electron--phonon interaction. Indeed, the
\qty{107.9}{\milli\eV} mode has a near-negligible HR factor of
\num{0.002}, while the \qty{139.7}{\milli\eV} mode shows limited
coupling, with an HR factor of \num{0.011}. This distinction is
illustrated in the spectral function of electron--phonon coupling
shown in the inset of Fig.~\ref{fig:luminescence_embedded},
where the participation of modes above \qty{65}{\milli\eV} is depicted
by Lorentzian-type sharp peaks.

In the inset of Fig.~\ref{fig:luminescence_embedded}, we present the
spectral density of the electron--phonon coupling during emission.
This function was calculated using Eq.~\eqref{eq:3}, where the
$\delta$-functions for phonon modes below \qty{65}{\milli\eV} were
replaced with Gaussians, each with a width of
$\sigma = \qty{0.9}{\milli\eV}$. For the localized modes above
\qty{65}{\milli\eV}, Lorentzians were employed, characterized by a
half-width at half-maximum (HWHM) of $\qty{0.8}{\milli\eV}$. Regarding
the electron--phonon interaction with both resonant and bulk-like
vibrational modes, the spectral function displays five distinct peaks
below \qty{65}{\milli\eV}. The peaks at \qtylist{20;42}{\milli\eV}
arise from the high density of bulk-like modes, while the remaining
three originate from the interactions with quasi-localized vibrational
resonances. The effective shapes of these collective vibrations are
presented in Appendix~\ref{app:resonances}.

Figure~\ref{fig:luminescence_embedded} presents the calculated
luminescence lineshape (blue), obtained using
Eqs.~\eqref{eq:lineshape} and \eqref{eq:opt_spectral_func_gen}, in
comparison with normalized experimental luminescence data (gray) from
Ref.~\cite{2021--Tajima--APE--FBEI}. The homogeneous broadening
parameter, $\gamma$, was set to $\qty{0.1}{\milli\eV}$. The horizontal
axis is shifted relative to the ZPL, similarly to
Fig.~\ref{fig:vibrational_structure_embedded}. The convergence of both
the spectral function and the luminescence lineshape with respect to
supercell size is discussed in detail in Appendix~\ref{app:embedding}.

The calculated emission lineshape in
Fig.~\ref{fig:luminescence_embedded} demonstrates remarkable agreement
with the experimental data. Various features in the theoretical
spectrum, arising from different modes, align well with the peaks and
dips observed in the experiment. The relative intensities of these
features are described accurately as well. The localized modes at
\qtylist{65.8;72.8;92.7}{\milli\eV} and the resonances at around
\qtylist{32;54;62}{\milli\eV} relative to the ZPL are clearly
replicated in the theoretical lineshape and correspond to pronounced
peaks in the experiment. The influence of bulk phonons is evident
through broad peaks at around \qtylist{20;42}{\milli\eV} from to the
ZPL. The total HR factor $S_{\mathrm{tot}}$ for the emission is
calculated to be \num{1.79}, while the Debye--Waller (DW) factor
$\omega_{\mathrm{ZPL}}$, which specifies the relative weight of the
intensity of the ZPL, is estimated to be about \qty{20}{\percent}.

While the overall agreement of the calculated emission spectrum with
the experimental data is outstanding, some discrepancies are found.
The main one is the absence of a sharp peak and its shoulder in the
calculated lineshape at around \qty{76}{\milli\eV} from the ZPL; the
origin of this feature remains an open question.

\begin{figure}
  \centering
  \includegraphics[width=\linewidth]{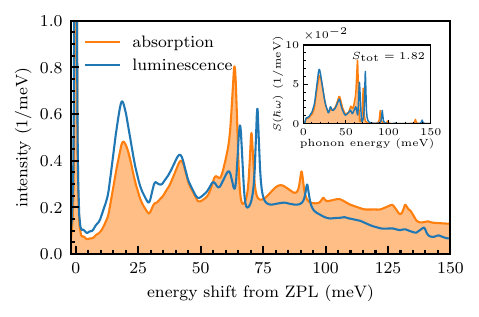}
  \caption{\label{fig:absorption_embedded} Calculated normalized
    absorption (orange) and luminescence (blue, from
    Fig.~\ref{fig:luminescence_embedded}) lineshapes of the C-center
    in silicon. Energies are taken relative to the ZPL value. The
    luminescence lineshape is inverted with respect to the ZPL for
    comparison. The inset figure contains the comparison between the
    spectral functions of electron--phonon coupling during absorption
    and luminescence. The results correspond to calculations for the
    $18 \times 18 \times 18$ supercell (\num{46656} atomic sites).}
\end{figure}

Figure~\ref{fig:absorption_embedded} presents the calculated
absorption lineshape (orange) compared to the computed luminescence
lineshape (blue) from Fig.~\ref{fig:luminescence_embedded}. The latter
is inverted with respect to the ZPL for comparison. The horizontal
axis is shifted relative to the ZPL. The inset figure compares the
spectral functions of the electron--phonon coupling during absorption
and luminescence. The calculated absorption lineshape corresponds to
the same system size and functional as in
Fig.~\ref{fig:luminescence_embedded}, but with the vibrational modes
of the excited state as opposed to those of the ground state in the
case of luminescence.

Upon comparing the calculated absorption and emission lineshapes,
along with the corresponding spectral functions presented in the
inset, it is evident that the spectra exhibit qualitative
similarities, particularly within approximately \qty{60}{\milli\eV} of
the ZPL. Beyond this threshold, the energies of the peaks in the
absorption lineshape, influenced by the localized modes, shift to
lower values compared to the corresponding peaks in the luminescence
lineshape. Notably, the localized modes at
\qtylist{70.3;90.4;99.0;131.8}{\milli\eV} are clearly involved in the
absorption process. The total HR factor $S_{\mathrm{tot}}$ for
absorption is calculated to be \num{1.82}, which is comparable to the
value of \num{1.79} for luminescence. This similarity indicates that
the vibrational structures of both the ground and excited states are
closely aligned regarding the phonons' forms and energies.

The disparity in the relative spectral weights of luminescence and
absorption lineshapes can be attributed to the $\omega^{\kappa}$
prefactor in Eq.~\eqref{eq:lineshape}. In the context of luminescence,
where $\kappa = 3$, this leads to a more pronounced increase in the
weight of the lineshape near the ZPL compared to absorption, which is
characterized by $\kappa = 1$. Consequently, this results in a
substantial asymmetry between the two spectra, as illustrated in
Fig.~\ref{fig:absorption_embedded}. Conversely, the asymmetry observed
in the electron--phonon coupling functions $\func{S}{\hbar\omega}$
(see inset of Fig.~\ref{fig:absorption_embedded}) is comparatively
less pronounced and arises purely from the differences in the
vibrational modes between the ground and excited states.

When analyzing the theoretical absorption lineshape of the C-center,
it is important to recognize that the actual absorption in bulk
silicon may differ from the calculated bound-to-bound transitions.
This discrepancy arises because the ionization threshold, which marks
the onset of electron promotion into the conduction band continuum,
lies only a few tens of \unit{\milli\eV} above the ZPL. Specifically,
Ref.~\cite{1985--Thonke--JPCSSP--CLDI} estimates this threshold to be
just \qty{38}{\milli\eV} above the ZPL. Consequently, the actual
absorption spectrum should reflect a convolution of both the
photoionization and the bound-to-bound absorption cross-sections.
Nevertheless, we anticipate that some of the sharp features in the
theoretical lineshape may still be observable in experimental
absorption measurements.

\begin{figure}
  \centering
  \includegraphics[width=\linewidth]{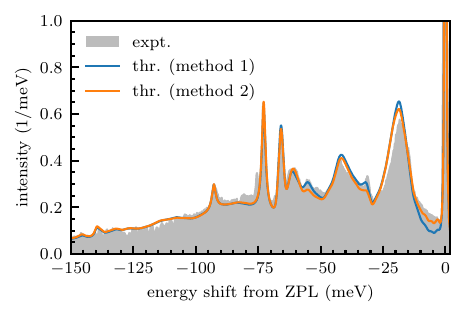}
  \caption{\label{fig:luminescence_embedded_comparison} Comparison of
    experimental (gray) and calculated normalized luminescence
    lineshape using two methodologies. The first methodology
    (method~1, from Fig.~\ref{fig:luminescence_embedded}) models the
    structural properties of the excited state using a positively
    charged defect in its ground state. The second methodology
    (method~2) uses the \dSCF\ approach, approximating the excited
    state as a triplet state $\ket{T}$. Experimental data are taken
    from Ref.~\cite{2021--Tajima--APE--FBEI}.}
\end{figure}

As previously discussed, the calculated lineshapes presented in this
study were obtained by modeling the structural properties of the
exciton-like excited singlet state $\ket{S_{1}}$ using the geometry of
the positively charged defect in its ground state. In
Fig.~\ref{fig:luminescence_embedded_comparison}, we compare the
luminescence lineshape from this approach (method~1) with that
obtained using the \dSCF\ methodology (method~2), in which the excited
triplet state $\ket{T}$ is used as an approximate model for the
excited singlet state $\ket{S_{1}}$. Additionally, these theoretical
lineshapes are compared with the experimental emission data from
Ref.~\cite{2021--Tajima--APE--FBEI}. Both theoretical approaches yield
almost identical results and show excellent agreement with the
experimental data. However, some discrepancies are observed between
the two methodologies. The first is in the treatment of acoustic
phonons near the ZPL, where method~2 provides a more accurate
description. The second discrepancy involves the resonances at
\qtylist{32;54}{\milli\eV} from the ZPL, where method~1 offers a
better description. A common feature of these methods is the failure
to describe the sharp peak and its shoulder at around
\qty{76}{\milli\eV} from the ZPL in the experimental data.

\section{Conclusions\label{sec:conclusion}}

In this work, we have calculated the thermodynamic properties of \Ci\
and \Oi\ interstitials, and the \CiOi\ complex, commonly referred to
as the C-center, achieving good agreement with the experimental data
and previous theoretical studies. By applying an advanced embedding
methodology, which enables the precise modeling of a defect's phonon
structure in the dilute limit, we have accurately characterized the
vibrational structure of the C-center in silicon. This approach offers
deep insights into the electron--phonon interactions of this defect
complex, uncovering a complicated interplay between localized and
bulk-like phonon modes. These modes, in turn, collectively influence
the intricate luminescence lineshape of the C-center, characterized by
many localized vibrational modes and resonances. Our calculated
theoretical lineshape exhibits outstanding correspondence with the
low-temperature experimental emission data, successfully reproducing
nearly all the features observed in the latter. This substantial
agreement further supports the assignment of the experimental
\qty{790}{\milli\eV} line to the \CiOi\ complex.

Furthermore, we demonstrated that the transitions involving single
exciton-like states can be effectively modeled by approximating the
geometry of the bound exciton state with that of a charged defect
while neglecting the contribution of the bound electron in the
hydrogenic orbital. This approximation simplifies the analysis of
optical lineshapes for such transitions without compromising accuracy.
Therefore, our work advances modern methodologies for precisely
describing the optical properties of defects exhibiting exciton-like
states within a first-principles computational framework. This
advancement enhances the ability to identify defects, particularly in
the search for those with potential applications in quantum
technologies.


\section*{Acknowledgements}

Financial support was kindly provided by the Research Council of
Norway and the University of Oslo through the frontier research
project QuTe (no. 325573, FriPro ToppForsk-program). Computational
resources were provided by the supercomputer GALAX of the Center for
Physical Sciences and Technology (FTMC), Vilnius, Lithuania, and the
High Performance Computing Center \enquote{HPC Saulėtekis} in the
Faculty of Physics, Vilnius University, Lithuania.


\clearpage
\bibliography{references}
\clearpage

\appendix
\section{Supercell embedding methodology and
  convergence\label{app:embedding}}
\renewcommand{\thefigure}{A\arabic{figure}}
\setcounter{figure}{0}

To accurately compute optical lineshapes with high precision and
resolution, it is necessary to go beyond the computational limitations
of directly accessible supercells, such as the $4 \times 4 \times 4$
supercell used in this study. Although such supercells are adequate
for capturing general features, they are insufficient for
high-accuracy requirements. To address this, we employ the embedding
methodology~\cite{2014--Alkauskas--NJP--FPTL,2021--Razinkovas--PRB--VVSI},
which allows the calculation of lattice relaxations and vibrational
modes in larger supercells where direct first-principles calculations
become computationally prohibitive. In our work, this method is
applied to $N \times N \times N$ supercells with
$N \in \set{8, 12, 16, 18}$.

The embedding approach relies on two key principles: (i) the
computation of vibrational spectra in large supercells and (ii)
lattice relaxation calculations for these supercells. The embedding
methodology is feasible because the interatomic forces in silicon are
short-ranged, and the forces induced by an electronic transition
rapidly diminish with increasing distance from the defect sites.
Accordingly, the Hessian matrix $\Phi_{\alpha,\beta}(m,n)$ [see
Eq.~\eqref{eq:Hessian}] for a large defect supercell is constructed
using a cutoff scheme. If atoms $m$ and $n$ are separated by a
distance greater than a predefined cutoff radius $r_{\mathrm{bulk}}$,
their corresponding Hessian elements are set to zero. We use the
Hessian elements from the defect supercell calculations for atom pairs
within the defect region, i.e., those separated from one of the defect
atoms by a distance less than the defect cutoff radius
$r_{\mathrm{defect}}$. Bulk values are used for the remaining atom
pairs. In this study we chose
$r_{\mathrm{bulk}} = \qty{10.855}{\angstrom}$ and
$r_{\mathrm{defect}} = \qty{8.314}{\angstrom}$. Finally, using
Eq.~\eqref{eq:5}, we estimate the relaxation component $\Delta Q_{k}$
along each vibrational mode $k$ by projecting the forces obtained
after the electronic transition within the geometry of the initial
state onto the vibrational modes of the embedded supercell.

\begin{figure}
  \centering
  \includegraphics[width=\linewidth]{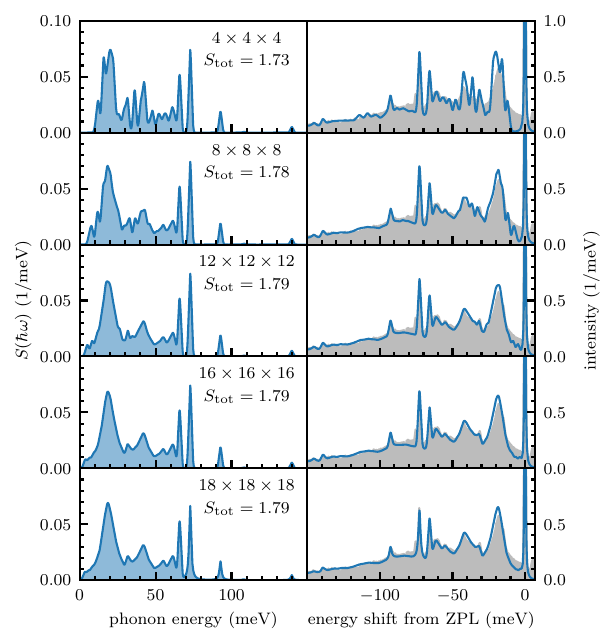}
  \caption{\label{fig:luminescence_embedded_convergence} Convergence
    of theoretical luminescence lineshapes with respect to supercell
    size. Supercells range in size from $4\times 4\times 4$ (\num{514}
    sites) to $18\times 18\times 18$ (\num{46658} sites). The
    lineshape of the $4\times 4\times 4$ supercell was computed
    without the embedding procedure.}
\end{figure}

Figure~\ref{fig:luminescence_embedded_convergence} presents the
convergence of the spectral density of electron--phonon coupling (left
column) and the luminescence lineshapes (right column) as a function
of supercell size, ranging from $4 \times 4 \times 4$ (\num{514}
atoms) to $18 \times 18 \times 18$ (\num{46658} atoms). These
calculations were performed using the SCAN functional. The spectral
density was computed using Eq.~\eqref{eq:3}, where the
$\delta$-functions for phonon modes below \qty{65}{\milli\eV} were
approximated by Gaussians with a width of
$\sigma = \qty{0.9}{\milli\eV}$. In contrast, the localized modes
above \qty{65}{\milli\eV} were modeled using Lorentzians with a HWHM
parameter of $\qty{0.8}{\milli\eV}$.
Fig.~\ref{fig:luminescence_embedded_convergence} demonstrates that,
for the chosen smoothing parameters, the convergence of the spectral
function and the optical lineshape is achieved with the
$16 \times 16 \times 16$ supercell, as the difference between the
results for the $16 \times 16 \times 16$ and $18 \times 18 \times 18$
supercells is minimal.

\section{Visualization of bulk-like and quasi-local vibrational
  modes\label{app:resonances}}
\renewcommand{\thefigure}{B\arabic{figure}}
\setcounter{figure}{0}

\begin{figure}
  \centering
  \includegraphics[width=\linewidth]{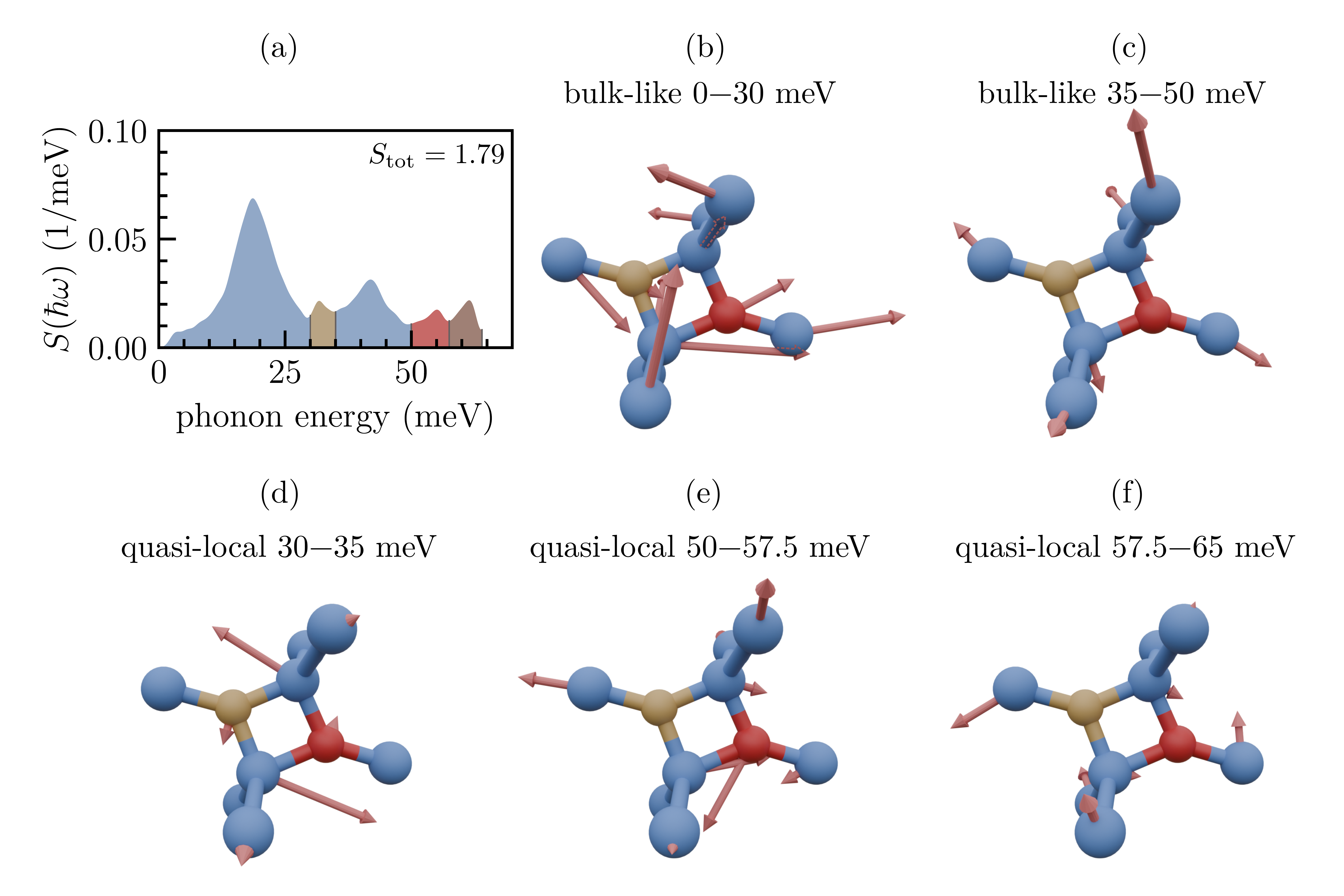}
  \caption{\label{fig:resonances} Effective shapes of collective
    vibrations. (a) Spectral density of electron--phonon coupling,
    indicating selected frequency regions. (b)--(c) Effective modes
    resulting from delocalized bulk-like vibrations. (d)--(f)
    Quasi-localized resonant modes. Note that the amplitude of
    resonant modes was chosen to be twice as large as that of
    bulk-like modes for improved visibility. Amplitude scaling was
    optimized for visual clarity.}
\end{figure}

To illustrate the electron--phonon interaction contributions from the
continuum of modes below \qty{65}{\milli\eV}, we computed the
effective vibrational modes using the following equation:
\begin{equation}
  \Delta R_{\mathrm{eff},m\alpha}
  = \sum_{k \in \Omega} \Delta Q_{k} \eta_{k; m\alpha} \sqrt{M_{m}}.
\end{equation}
Here $\Delta Q_k$ is obtained from Eq.~\eqref{eq:delta_q}, and
$\Omega$ represents the selected frequency interval encompassing a set
of phonons. For example, to represent bulk-like vibrational modes that
arise from high phonon density regions, we selected frequency
intervals of \qtyrange{0}{30}{\milli\eV} and
\qtyrange{35}{50}{\milli\eV}. Figure~\ref{fig:resonances}(a) displays
the spectral density of the electron--phonon coupling, highlighting
the chosen frequency regions below \qty{65}{\milli\eV}.
Figures~\ref{fig:resonances}(b) and (c) show the effective vibrational
modes localized on defect atoms and their adjacent neighbors,
corresponding to bulk-like vibrations. In contrast,
Figs.~\ref{fig:resonances}(d)--(f) depict quasi-localized vibrational
resonances, computed for the frequency ranges
\qtyrange{30}{35}{\milli\eV}, \qtyrange{50}{57.5}{\milli\eV}, and
\qtyrange{57.5}{65}{\milli\eV}, respectively.

\end{document}